\documentclass[serif,twocolumn,alpha-refs]{wiley-article}
\paperdoi{10.1002/asna.202113881}

\usepackage{siunitx}
\usepackage{setspace}
\usepackage{hyperref}
\papertype{latex template}
\paperfield{Based on Astronomische Nachrichten}
\title{Implementation of two-field inflation for cosmic linear anisotropy solving system}

\author[1]{Braulio Morales-Mart\'{\i}nez}
\author[1,2]{Gustavo Arciniega}
\author[3]{ Luisa G. Jaime}
\author[2]{Gabriella Piccinelli}

\affil[1]{Departamento de F\'{\i}sica, Facultad de Ciencias, Universidad Nacional Aut\'onoma de M\'exico, Ciudad de M\'exico, M\'exico}
\affil[2]{Centro Tecnol\'ogico, Facultad de Estudios Superiores Arag\'on, Universidad Nacional Aut\'onoma de M\'exico, Ciudad de M\'exico, M\'exico}
\affil[3]{Departamento de F\'{\i}sica, Instituto Nacional de Investigaciones Nucleares, Ciudad de M\'exico, M\'exico}

\corraddress{Braulio Morales-Mart\'inez, Departamento de F\'{\i}sica, Facultad de Ciencias, Universidad Nacional Aut\'onoma de M\'exico, A.P. 50-542, Ciudad de M\'exico 04510, M\'exico}
\corremail{jorgebraulio@ciencias.unam.mx}

\runningauthor{Braulio Morales-Mart\'{\i}nez et al.}

\begin{document}
\setstretch{0.8}
\begin{frontmatter}
\maketitle

\begin{abstract}
We outline the modifications in the numerical Boltzmann code Cosmic Linear Anisotropy Solving System (CLASS) in order to include extra inflationary fields. The functioning of the code is first described, how and where modifications are
meant to be done are later explained.  In the present study, we focus on the modifications needed for the implementation of a two-field inflationary model, with canonical kinetic terms and a polynomial potential with no cross terms, presenting preliminary results for the effect of the second field on the spectra. The adaptability of the code is exploited, making use of the classes and structures of C and the generic Runge-Kutta integration tool provided by the program.

\end{abstract}
\end{frontmatter}
\section{Introduction}
Inflation is an early period of accelerated expansion of the universe. It successfully explains the flatness and horizon problems and also provides a mechanism for the generation and evolution of density perturbations that give shape to the universe we observe today \citep{weinberg2008cosmology}. There are many ways to explain the acceleration of this period; the most popular being the introduction of one scalar field, called the inflaton. In an homogeneous universe, this scalar field depends on time alone and has an associated potential energy function determined for each inflation theory.
However, there is no reason to have a single field driving inflation, multiple-fields scenarios are equally valid particularly when considering high-energy theories used to describe inflation, such as supersymmetry theories and string theory, which have multiple degrees of freedom and contain several fields that could participate in inflationary dynamics \citep{Nilles:1983ge}.

During inflation, quantum vacuum fluctuations of the fields previously mentioned are stretched due to the accelerated expansion of space and become classical perturbations when they leave the Hubble horizon. Once the inflation period has ended,
these perturbations will be the seed for the temperature anisotropies observed in the cosmic microwave background (CMB) and the
inhomogeneous distribution of galaxies at large scales \citep{Gong_2017}.

There are not many multifield cases which can be solved analytically to obtain the perturbations and thus make predictions to be compared with the CMB observations. One option is to use approximation schemes such as the Slow-Roll approximation.

On the other hand, numerical codes, such as Cosmic Linear Anisotropy Solving System (CLASS), are a powerful option to explore a wide variety of cosmological model predictions \citep{Blas_2011}. As far as we know, CLASS does not support models involving multifield inflation, and even though there are options that numerically solve multifield inflation, like MULTIMODECODE \citep{Price_2015}, they provide only solutions for the inflationary epoch and do not give a complete cosmological model view, and there is no simple way to implement this tool in CLASS.

In this paper, we describe the changes needed in the primordial module of CLASS, so it can include two-field inflationary models with canonical kinetic terms \citep{Lalak:2007vi}, with the aim that this modification will open the path to work numerically with more complicated models with multiple fields and noncanonical kinetic terms.
\section{Two-Field Inflation}
The general action for multifield inflation minimally coupled to gravity is given by
\begin{equation}\label{eq:action}
    S = \int d^4 x \sqrt{-g} \left( \frac{m_{Pl}^2}{2}R - \frac{1}{2} G_{pq} g^{\mu \nu} \partial _{\mu} \phi ^p  \partial _{\nu} \phi ^q - V(\phi) \right),
\end{equation} 
where $m_{Pl}$ is the Planck mass, $G_{pq}$ is the metric of the field space, and $V$ the potential as a function of the fields.
From this action, assuming $G_{pq}=\delta _{pq},$ that is, with canonical kinetic terms, the background dynamics can be obtained, resulting in:
\begin{equation}\label{frid1}
    H^2 = \frac{1}{3m_{pl}^2} \left[\frac{1}{2}\dot{\phi }^2 + V(\phi ) \right],
\end{equation}
\begin{equation}\label{field1}
\ddot{\phi _p} +3H\dot{\phi _p} + \frac{\partial V(\phi)}{\partial \phi ^p } = 0,
\end{equation}
where $H$ is the Hubble parameter, the dot indicates the derivatives with respect to the cosmic time $t$,  and $\dot{\phi }^2\equiv G_{pq}\dot{\phi ^p}\dot{\phi ^q}$ \citep{thesis}.

Rewriting equations (\ref{frid1}) and (\ref{field1}) in conformal time, we have:
\begin{equation}\label{eq:b1}
    \mathcal{H} ^2=  \dfrac{1}{3 m_{pl}^2} \left( \dfrac{1}{2}  {\phi '}^2 + \dfrac{1}{2}  {\chi '}^2+ a^2  V(\phi , \chi ) \right),
\end{equation}
\begin{equation}\label{eq:b2}
    \phi '' +2a\mathcal{H} \phi' + a^2 \dfrac{\partial V}{\partial \phi}=0,
\end{equation}
\begin{equation}\label{eq:b3}
    \chi '' +2a\mathcal{H} \chi' + a^2 \dfrac{\partial V}{\partial \chi}=0,
\end{equation}
where $\mathcal{H} =\left( a'/a \right) $, the primes are derivatives with respect to the conformal time $\tau $, $\phi$ and $\chi$ the two fields, and $V$ is their potential. For our trial model \citep{Lalak:2007vi}\begin{equation}
V(\phi, \chi)=\frac{1}{2}(m_\phi \phi^2+m_\chi \chi^2).
\end{equation}

According to  \citep{Brandenberger:2003vk}, the scalar perturbations can be expressed in terms of the \textit{Mukhanov variable} for each field, a la single field inflation:
\begin{equation}\label{eq:mukhanov}
u^p=a\left(\delta \phi^p + \dfrac{{\phi _0}^p{}'}{\mathcal{H}}  \Psi \right),
\end{equation}
with $\Psi$ being the metric perturbation potential in the Newtonian gauge and $\phi _0^p$ the background part of the p-th field.

The equations of motion for this variable can be obtained through the variation of equation \eqref{eq:action} for each field, resulting in:
\begin{equation}\label{eq:pertur}
    u^p_k{}'' -k^2 u^p_k - \dfrac{z_p{}''}{z_p} u^p_k = 0,
\end{equation}
where $z_p = a(\phi _0^p{}')/(\mathcal{H})$, and the equation is in Fourier space, with $k$ the wave number \citep{Brandenberger:2003vk}.
\section{CLASS: Modules and classes}
CLASS works by making use of classes in the Object Oriented Programming (OOP) sense. It has several modules, each one having functions and variables, some of which are ``public'' and others ``private''. Each module has public functions designed for other modules to use them in order to access the quantities computed by said module.
\subsection{Primordial}
The module \textit{"primordial"} is the one in charge of solving the inflationary process. Its main purpose is to compute the Primordial Spectra. In order to do this, the code has two options: (a) to receive the parameters describing the analytical shape of the spectra and (b) to compute the spectra in a numerical way by receiving the expression for the potential in terms of the inflaton field ($V(\phi)$), the value of the potential at the end of inflation, or the Hubble parameter function in terms of the inflaton field. For each of the numerical options, canonical kinetic terms are considered.

We will describe the numerical way in which this module computes the primordial spectra.
\subsection{Variables}
The required variables for this module are the background and perturbation quantities, whose indices, in a general vector of quantities, are handled through the structure \texttt{"primordial struct"}, defined in the 
\texttt{"primordial.h"} file (structure in the context of OOP).

\subsubsection{Background} $\{a$, $\dot{a}$, $\phi$, $\dot{\phi} $, $\ddot{\phi}\}$.
Here, the derivatives correlate to the cosmic time, but the program has the option to work with both cosmic and conformal time. 
If the value of the potential $V$ at the end of inflation is given, then the quantity $\ddot{\phi}$ can be skipped by the program.

\subsubsection{Perturbations} $\{Re(u_k)$, $Im(u_k)$, $Re(u_k')$, $Im(u_k')$  $Im(\mu _k')\}$.
The scalar perturbations in the code are handled in terms of the Mukhanov variable introduced in \eqref{eq:mukhanov}.

In the Fourier space, we will have complex quantities whose real and imaginary parts will be solved separately by the program. The perturbations are given in conformal time through the program.

Each of this quantities corresponds to one variable in the code and will evolve via the field and perturbation equations.
\vspace{-.33 cm}
\subsection{Evolution}
As we have mentioned, the main purpose of the primordial module is to compute the Primordial Spectra value for a (finite) set of wavenumbers $k$ within a certain range. When this is performed, an interpolation process is performed in order to compute any value of $k$ between the specified range. 
The main function of this module is \texttt{"primordial\_inflation\_solve\_inflation"}. Here, to check if evolution is suitable, the background quantities are numerically evolved a bit forward and a bit backward from a pivot value of the inflaton field at which a typical observable wavelength \texttt{k\_\text{pivot}} (chosen to be roughly in the middle of the range probed by CMB and large scale structure experiments) crosses the Hubble radius during inflation, that is, \texttt{k\_\text{pivot} = a\_\text{pivot}*H\_\text{pivot}}.
Once this pivot value of the inflaton is found, initial conditions for the inflaton field must be set. This is done by calling the numerical routine designated to find the attractor solution for $\phi$ in the phase-space, and then, the value of $(\dot{\phi})_{\text{pivot}}$ is set to match the value of the attractor solution \citep{evolucion}.

After this, the routine computes the perturbations and stores the results of the primordial spectra for each value of $k$. This is done in a loop over $i$, integrating the background quantities from an initial given time $t$ to a final time $t+\Delta _i t$ (\textit{i.e.}, the step is adaptive for each i-th moment of the loop) and checking the conditions of crossing the Hubble Horizon and the stabilization of the curvature. If these conditions are met, the loop stops, and the final values of tensor perturbation and curvature are stored to compute the primordial spectrum at this $k$ \citep{classmanual}.
\section{Modifying the program}
As mentioned before, inflation models that take into account several fields are relevant for various reasons.

Our first approach to deal with many fields will be to add a feature in CLASS for solving the most simple of the two-field inflation models: double inflation with canonical kinetic terms.
\subsection{Adding new variables to the code: Potential Terms and an extra Field}
The first step of allowing CLASS to work with more fields is to make new variables for the new field and its derivatives. We will also need more parameters describing the shape of the potential. 
Currently, only a polynomial function with no cross-terms up to 4th grade is considered in CLASS, i.e. (from here on, the fields background part $\phi _0$ and $\chi _0$ will be denoted simply as $\phi$ and $\chi$, respectively)
\begin{eqnarray}\nonumber
V(\phi, \chi) &=&  V_{00} +  V_{01}\phi +V_{02}\phi ^2 + V_{03}\phi ^3 + V_{04}\phi ^4
+ \\
&& V_{10} +  V_{11}\chi +V_{12}\chi ^2 + V_{13}\chi ^3 + V_{14}\chi ^4.
\end{eqnarray}
To do so, we have to add a new member variable to the primordial structure in the \texttt{primordial.h} file.

In this structure, each quantity has an integer variable containing its index; this index is passed as an argument to the vector of quantities and the vector of its derivatives to access the values of this quantities.

Keeping in mind that our future goal is to make a multifield modification, our strategy was to replace the integer variables \texttt{index$\_$in$\_$phi} and \texttt{index$\_$in$\_$dphi}, each one for a vector of integers, creating empty pointers that will be allocated later on with the size determined by a new integer variable, \texttt{n\_of\_fields}, that will be read from the user input. For now, our case is \texttt{n\_of\_fields} = $\{1 , 2\}$.
 These pointers will be allocated using the global allocating functions of CLASS to obtain a vector of integers, with the i-th entry of the vector being the index of the i-th inflation field.
For the potential parameters, we will create an array variable,  \texttt{V\_multi\_field}, with dimensions (\texttt{5 x n\_of\_fields}), again declared in the \texttt{primordial.h} file, and will use a double pointer of the double type.

This pointer will be allocated in the same way as the previous ones and the array will be filled with a double loop.

\subsection{Modifying the field equations}
The differential equations describing the background quantities (the scale factor $a$ and the two fields $\phi$ and $\chi$) will be solved with the generic Runge-Kutta integrator provided by CLASS (which is already designed to deal with systems of differential equations with  multiple variables), and we will have only one extra equation for the new field $\chi$, exactly like in equations \eqref{eq:b1}-\eqref{eq:b3}.

This equations will be provided to the Runge-Kutta method to solve for $a$, $\phi$, $\dot{\phi}$, $\chi$, $\dot{\chi}$.

\subsection{Perturbations for two fields}
We are currently working on the numerical process to deal with the perturbations, both scalar and tensor, of two or more fields.
The scalar perturbations are in terms of the Mukhanov variable for each field (we add extra variables in form of a vector [pointers] in the same way as we did for the fields), with the equation of motion given by \eqref{eq:pertur}.

For each $k$, these equations, which are dependent on the evolution of the fields through the terms $z_{\phi}$ and $z_{\chi}$, will be solved numerically, with the same numerical integrator that solves the evolution of the background quantities and in a similar step-by-step loop until $k >> aH$ and the curvature is stable.
\subsection{Results for Two Field Inflation in CLASS}
Using these modifications, we have been able to implement the second field and the corresponding background equations for the process leading to the primordial fluctuation spectra.
\begin{figure}
\includegraphics[width=70mm]{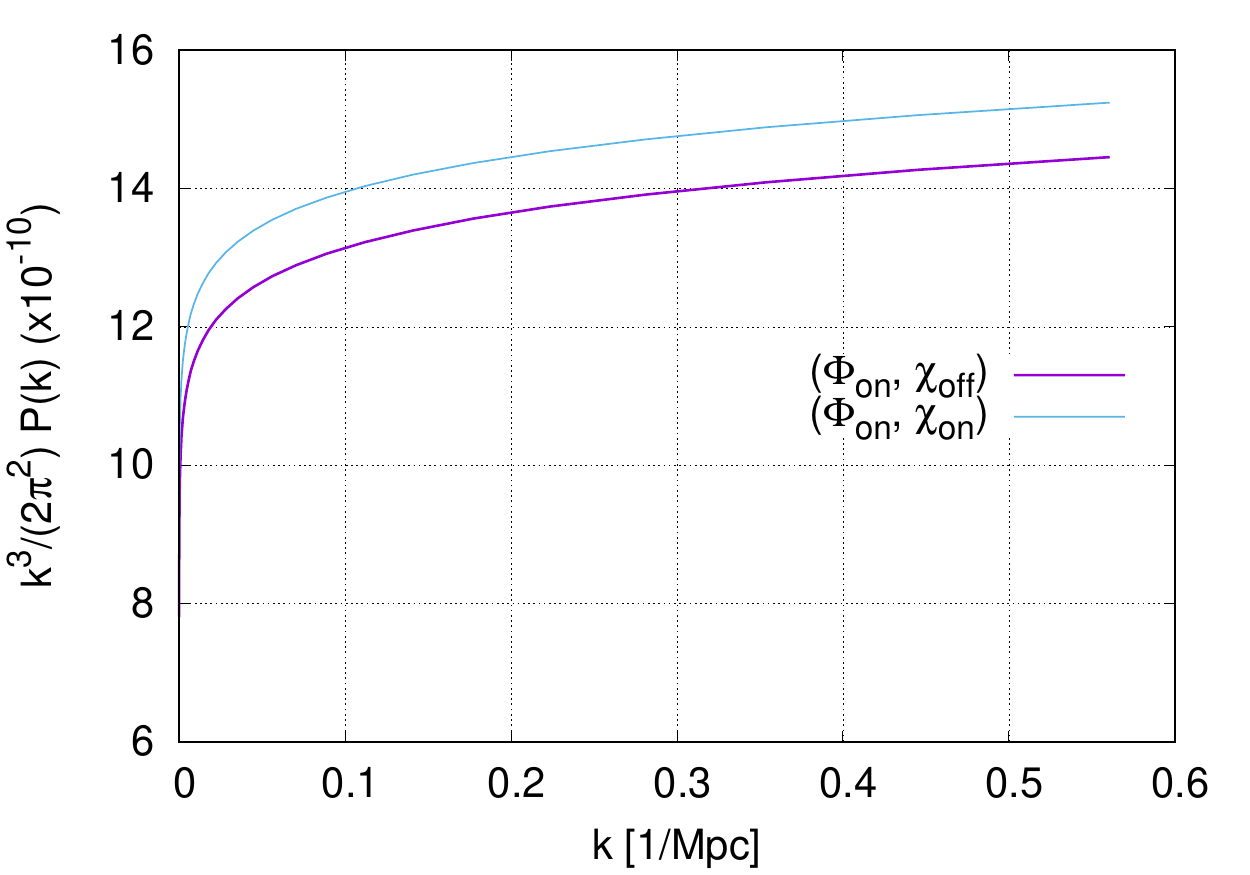} 
\caption{\small{Primordial spectrum $P(k)$ for: $(a)$ the standard field $\phi$ on and the second off ($\phi_{\text{on}}$, $\chi_{\text{off}}$) (Purple) and $(b)$ two fields on ($\phi_{\text{on}}$, $\chi_{\text{on}}$) (Blue).}} \label{fig1}
\end{figure}
\begin{figure}
\includegraphics[width=70mm]{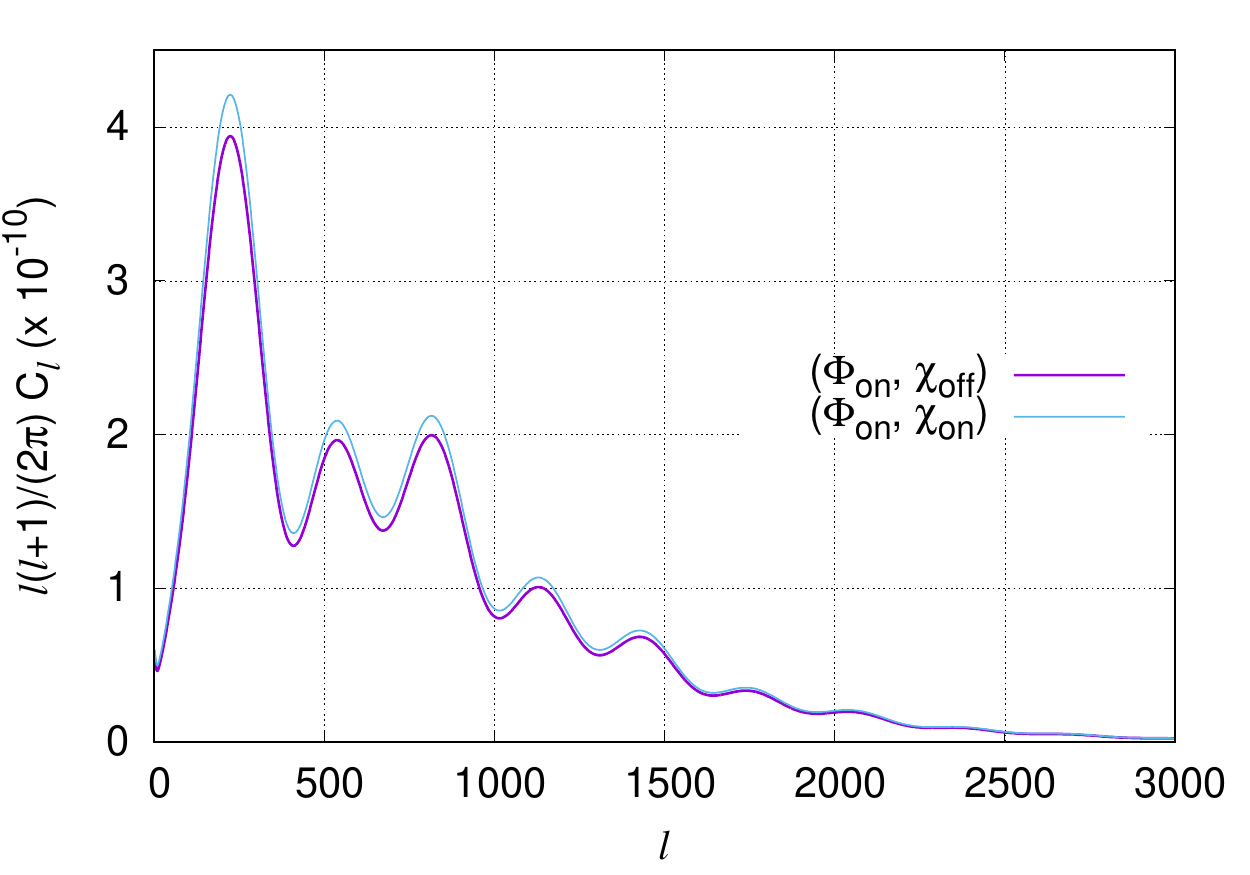} 
\caption{\small{Temperature Spectrum for two cases:  $(a)$ the standard field $\phi$ on and the second off ($\phi_{\text{on}}$, $\chi_{\text{off}}$) (Purple) and $(b)$ two fields on ($\phi_{\text{on}}$, $\chi_{\text{on}}$) (Blue). Notice that the modifications we have made to the code with the addition of the second field have an impact in the spectrum.}} \label{fig1}
\end{figure}
In figures 1 and 2, we compare the Primordial and Temperature spectra for a single field and two fields in the background, fluctuating only the principal field $\phi$ in both cases, considering for now that the fluctuation of the second field is negligible. 
In these figures,  the potential parameters used for the field $\phi$ are: $V_{00}=1\times 10^{-13}$, $ V_{01}=-1\times 10^{-14}$, and $V_{02}=7 \times 10^{-14}$ as in the default CLASS input parameters, and for the field $\chi$, we set the same parameters multiplied by a factor of $10^{-2}$.
\section{Conclusions and Perspectives}
We presented the primordial and temperature spectra for the case of single- and double-field inflation in the background, with fluctuations only in the main field $\phi$ in both cases.
The modifications that we outlined in this work will allow CLASS to work its entire routine, resulting in a prediction of the CMB spectrum and quantities related to a two-field inflationary model (with canonical kinetic terms and a polynomial potential). As a result of the flexible handling of quantities through the use of classes and structures in C and generalized processes to numerically solve physical equations in the software, extra quantities for background and scalar perturbations are easily introduced without disrupting or requiring major changes in the organization and tools of the code.

Our modification could be implemented in extended numerical codes like $hi\_class$ \citep{Zumalacarregui:2016pph}, where alternative models involving scalar fields in the action are studied.
The tensor and scalar perturbations are yet to be numerically implemented for the second field. Once this is done, the possibility to allow CLASS to work with multifield models with noncanonical kinetical terms will be the natural extension of this work. 

\section*{Acknowledgments}

 The authors acknowledge the support from PAPIIT IN120620 and thank the higher-order gravity research group, particularly C. Benitez, for fruitful discussions. BMM acknowledges the fellowship from PAPIIT IN120620. GA acknowledges the postdoctoral fellowship from DGAPA-UNAM. LGJ acknowledges the financial support from SNI (CONACyT) and Instituto Nacional de Investigaciones Nucleares (ININ).
\nocite{*}
\bibliography{sample}

\begin{thebibliography}{11}
\expandafter\ifx\csname natexlab\endcsname\relax\def\natexlab#1{#1}\fi
\expandafter\ifx\csname url\endcsname\relax
  \def\url#1{\texttt{#1}}\fi
\expandafter\ifx\csname urlprefix\endcsname\relax\def\urlprefix{URL: }\fi

\bibitem[{Blas et~al.(2011)BlasLesgourgues and Tram}]{Blas_2011}
Blas, D.Lesgourgues, J. and Tram, T. (2011) The cosmic linear anisotropy
  solving system (class). part ii: Approximation schemes.
\newblock \textit{JCAP} \textbf{2011} 034–034.

\bibitem[{Brandenberger(2004)}]{Brandenberger:2003vk}
Brandenberger, R.~H. (2004) {Lectures on the theory of cosmological
  perturbations}.
\newblock \textit{Lect. Notes Phys.} \textbf{646} 127--167.

\bibitem[{Gong(2017)}]{Gong_2017}
Gong, J.-O. (2017) Multi-field inflation and cosmological perturbations.
\newblock \textit{IJMP D} \textbf{26} 1740003.

\bibitem[{Lalak et~al.(2007)LalakLangloisPokorski and Turzynski}]{Lalak:2007vi}
Lalak, Z.Langlois, D.Pokorski, S. and Turzynski, K. (2007) {Curvature and
  isocurvature perturbations in two-field inflation}.
\newblock \textit{JCAP} \textbf{07} 014.

\bibitem[{Lesgourgues(2006)}]{evolucion}
Lesgourgues, J. (2006) Lecture notes.
\newblock \urlprefix\url{https://
  lesgourg.github.io/courses/Inflation_EPFL.pdf}.

\bibitem[{Lesgourgues and Hooper()}]{classmanual}
Lesgourgues, J. and Hooper, D. () \textit{CLASS MANUAL}.

\bibitem[{Nilles(1984)}]{Nilles:1983ge}
Nilles, H.~P. (1984) Supersymmetry, supergravity and particle physics.
\newblock \textit{Phys. Rept.} \textbf{110} 1--162.

\bibitem[{Price et~al.(2015)PriceFrazerXuPeiris and Easther}]{Price_2015}
Price, L.~C.Frazer, J.Xu, J.Peiris, H.~V. and Easther, R. (2015) Multimodecode:
  an efficient numerical solver for multifield inflation.
\newblock \textit{JCAP} \textbf{2015} 005–005.

\bibitem[{Weinberg(2008)}]{weinberg2008cosmology}
Weinberg, S. (2008) \textit{Cosmology}.
\newblock Cosmology. OUP Oxford.

\bibitem[{de~Wild(2018)}]{thesis}
de~Wild, T. (2018) \textit{Primordial Non-Gaussianity in the Single and
  Multi-Field Inflationary Scenarios}.
\newblock {B.S. Thesis} University of Groningen, Faculty of Science and
  Engineering.

\bibitem[{Zumalac\'arregui
  et~al.(2017)Zumalac\'arreguiBelliniSawickiLesgourgues and
  Ferreira}]{Zumalacarregui:2016pph}
Zumalac\'arregui, M.Bellini, E.Sawicki, I.Lesgourgues, J. and Ferreira, P.~G.
  (2017) {hi\_class: Horndeski in the Cosmic Linear Anisotropy Solving System}.
\newblock \textit{JCAP} \textbf{08} 019.

\end{thebibliography}
\section*{AUTHOR BIOGRAPHY}
\begin{biography}
B\textbf{raulio Morales Mart\'inez} is a physics undergraduate at UNAM, born in Guanajuato, M\'exico. He participated in several courses and seminars at the Mathematics Research Center (CIMAT by its Spanish acronym),
such as XIV Calculus Problem Solving Workshop, and
various chapters of Science Clubs Mexico (CdeCMx).
In 2020 he participated at the International Workshop
on Astronomy and Relativistic Astrophysics.
\end{biography}
\otherinfo{Morales-Mart\'inez B,
Arciniega G, Jaime LG, Piccinelli G.\\
Implementation of two-field inflation for cosmic
linear anisotropy solving system.\textit{Astron.}\textit{Nachr.}
2021;342:58–62. \href{https://doi.org/10.1002/asna.202113881}{https://doi.org/10.1002/asna.202113881}}
\end{document}